\documentclass[twocolumn,showpacs,amsmath,amssymb,pra,superscriptaddress]{revtex4-1}
\usepackage[utf8]{inputenc}
\usepackage{graphicx}
\usepackage{epstopdf}
\usepackage{epsfig}
\usepackage{dcolumn}
\usepackage{color}
\usepackage{bm}
\usepackage{mathrsfs}
\usepackage{bbold}
\usepackage{amssymb}
\usepackage{upgreek}

\begin{document}

\title{Quantum Uncertainty Dynamics}

\author{Md.~Manirul Ali}
\email{manirul@citchennai.net}
\affiliation{Centre for Quantum Science and Technology, Chennai Institute of Technology, Chennai 600069, India}

\date{\today}

\begin{abstract}
Quantum uncertainty relations have deep-rooted significance on the formalism of quantum mechanics.
Heisenberg's uncertainty relations attracted a renewed interest for its applications in quantum information science.
Robertson derived a general form of Heisenberg's uncertainty relations for a pair of arbitrary observables represented
by Hermitian operators. In the present work, we discover a temporal version of the Heisenberg-Robertson uncertainty
relations for the measurement of two observables at two different times, where the dynamical uncertainties crucially
depend on the time evolution of the observables. The uncertainties not only depend on the choice of observables, but
they also depend on the times at which the physical observables are measured. The time correlated two-time commutator
dictates the trade-off between the dynamical uncertainties. We demonstrate the dynamics of these uncertainty relations for a
spin-1/2 system and for a quantum harmonic oscillator. The temporal uncertainty relations discovered in this work can
be experimentally verified with the present quantum technology.
\end{abstract}

\maketitle

In classical mechanics, particles can have definite trajectory with precise position and momentum
at any given time. This picture is not valid in quantum mechanics. Heisenberg's uncertainty principle
gives a restriction on the accuracy of position and momentum measurements for quantum particles.
Heisenberg’s uncertainty principle \cite{Heisenberg,Kennard} states that if the momentum of a particle is
measured with an uncertainty $\Delta p$ then the position measurement uncertainty $\Delta x$ is restricted
by the relation
\begin{eqnarray}
\label{hup}
\Delta x \Delta p \ge \frac{\hbar}{2},
\end{eqnarray}
where $\hbar$ is the reduced Planck constant. Heisenberg in his 1927 paper \cite{Heisenberg} asserted
that this relation is a consequence of the quantum commutation rule for the position and momentum
operators $\left[x,p\right]=i\hbar$. Apart from position and momentum, uncertainty principle applies
to many other pairs of observables including phase and excitation number of a harmonic oscillator,
angle and the orbital angular momentum of a particle, and components of spin or total angular momentum.
Robertson derived \cite{Robertson} a general form of Heisenberg's uncertainty relations for a pair of arbitrary
observables represented by two Hermitian operators $A$ and $B$ as follows
\begin{eqnarray}
\label{Robert}
\Delta A \Delta B \ge \frac{1}{2} \Big| \langle [A,B] \rangle \Big|,
\end{eqnarray}
which gives a lower bound to the product of variances of the two observables. The expectation value and
the variances are obtained with respect to some state. The uncertainty relation in terms of sum of variances was
also derived for incompatible observables \cite{ArunPati}. Quantum uncertainty relations have deep-rooted
significance on the formalism of quantum mechanics \cite{Busch}. Uncertainty principle has been the focus
due to its applications in quantum information science \cite{RMP17}, specifically in entanglement detection
\cite{Entang1,Entang2}, security analysis in quantum cryptography \cite{Crypto}, and determining quantum
nonlocality \cite{Nonlocality}.

In the present work, we generalize Heisenberg-Robertson uncertainty relation (\ref{Robert}) for the
measurement of physical observables $A$ and $B$ at two different times $t_1$ and $t_2$. In this situation,
we derive the following uncertainty relation
\begin{eqnarray}
\label{hupT}
\Delta A(t_1) \Delta B(t_2) \ge \frac{1}{2} \Big| \langle [A(t_1),B(t_2)] \rangle \Big|,
\end{eqnarray}
where the variances $(\Delta A(t_1))^2 = \langle A^2(t_1) \rangle-\langle A(t_1) \rangle^2$ and
$(\Delta B(t_2))^2 = \langle B^2(t_2) \rangle-\langle B(t_2) \rangle^2$ measure the time-dependent
uncertainties for the observables $A(t_1)$ and $B(t_2)$ with respect to some state $| \Psi \rangle$.
The above uncertainty relation (\ref{hupT}) is a temporal version of the Heisenberg-Robertson uncertainty
relation (\ref{Robert}). The dynamical uncertainties $\Delta A(t_1)$ and $\Delta B(t_2)$ crucially depend
on the dynamics of the observables, where the time dynamics of the commutator $[A(t_1),B(t_2)]$
determines the trade-off between the measurement knowledge for the two observables. Next, we derive
the dynamical uncertainty relation (\ref{hupT}) when a quantum system evolves under a particular
Hamiltonian $H$. We consider here a time-independent Hamiltonian for simplicity. The time evolution of the
operators are governed by the Heisenberg equation of motion
\begin{eqnarray}
\label{At}
\frac{d}{dt} A(t) = \frac{1}{i\hbar} \left[ A(t),  H \right],
\end{eqnarray}
\vskip -0.5cm
\begin{eqnarray}
\label{Bt}
\frac{d}{dt} B(t) = \frac{1}{i\hbar} \left[ B(t),  H \right],
\end{eqnarray}
where $A(t)$$=e^{i H t / \hbar}$$A$$e^{ - i H t / \hbar}$ and $B(t)$$=e^{i H t / \hbar}$$B$$e^{ - i H t / \hbar}$ are
the Heisenberg picture operators. The operators $A(t)$ and $B(t)$ are Hermitian at any given time. We consider the
measurement of observables $A(t_1)$ and $B(t_2)$ at two different times $t_1$ and $t_2$. The expectation values
of the two Hermitian operators $A(t_1)$ and $B(t_2)$ with respect to a quantum state $|\Psi\rangle$ are
$\langle A(t_1) \rangle = \langle \Psi | A(t_1) | \Psi \rangle$ and $\langle B(t_2) \rangle = \langle \Psi | B(t_2) | \Psi \rangle$.
We define the fluctuation operators $\delta A(t_1)$ and $\delta B(t_2)$ as follows
\begin{eqnarray}
\label{dA}
&\delta A(t_1) = A(t_1) - \langle A(t_1) \rangle, \\
\label{dB}
&\delta B(t_2) = B(t_2) - \langle B(t_2) \rangle.
\end{eqnarray}
Then the following expectation values
\begin{eqnarray}
\label{fA}
\langle (\delta A(t_1))^2 \rangle = \langle \Psi | (\delta A(t_1))^2 | \Psi \rangle = (\Delta A(t_1))^2, \\
\label{fB}
\langle (\delta B(t_2))^2 \rangle = \langle \Psi | (\delta B(t_2))^2 | \Psi \rangle = (\Delta B(t_2))^2,
\end{eqnarray}
estimate the the mean square deviations or the variances of the two operators. The action of the fluctuation
operators (\ref{dA}) and (\ref{dB}) on the state $| \Psi \rangle$ are as follows
\begin{eqnarray}
\label{Psi1}
| \Psi_1 \rangle = \delta A(t_1) | \Psi \rangle = \left( A(t_1) - \langle A(t_1) \rangle \right) | \Psi \rangle,\\
\label{Psi2}
| \Psi_2 \rangle = \delta B(t_2) | \Psi \rangle = \left( B(t_2) - \langle B(t_2) \rangle \right) | \Psi \rangle.
\end{eqnarray}
The inner products of the states $| \Psi_1 \rangle$ and $| \Psi_2 \rangle$ satisfy the Schwarz inequality given by
\begin{eqnarray}
\label{Sch}
\langle \Psi_1 | \Psi_1 \rangle \langle \Psi_2 | \Psi_2 \rangle \ge \Big| \langle \Psi_1 | \Psi_2 \rangle \Big|^2.
\end{eqnarray}
Substituting $| \Psi_1 \rangle$ and $| \Psi_2 \rangle$ from Eqs.~(\ref{Psi1}) and (\ref{Psi2}) to the above
Schwarz inequality (\ref{Sch}) we have
\begin{eqnarray}
\label{Sch2}
\langle (\delta A(t_1))^2 \rangle ~ \langle (\delta B(t_2))^2 \rangle \ge \Big| \langle \delta A(t_1) \delta B(t_2) \rangle \Big|^2,
\end{eqnarray}
where all the expectation values in (\ref{Sch2}) are with respect to the state $| \Psi \rangle$. The expectation value
$\langle \delta A(t_1) \delta B(t_2) \rangle$ on the right of (\ref{Sch2}) can be expressed as an averaged sum of a commutator
$\langle [A(t_1), B(t_2)] \rangle/2$ and an anticommutator $\langle \{ \delta A(t_1), \delta B(t_2) \}\rangle/2$. Note that
$\delta A(t_1)$ and $\delta B(t_2)$ are Hermitian as the operators $A(t_1)$ and $B(t_2)$ are both Hermitian. Now the commutator
$[A(t_1), B(t_2)]$ is anti-Hermitian, the expectation value of which is pure imaginary. Whereas the anticommutator
$\{ \delta A(t_1), \delta B(t_2) \}$ is Hermitian and its expectation value is real. Hence the squared modulus
$\big| \langle \delta A(t_1) \delta B(t_2) \rangle \big|^2$ on the right hand side of (\ref{Sch2}) can be written as a sum of two terms
\begin{eqnarray}
\label{SqMod}
\frac{1}{4} \Big| \langle [A(t_1), B(t_2)] \rangle \Big|^2 + \frac{1}{4} \Big| \langle \{ \delta A(t_1), \delta B(t_2) \}\rangle \Big|^2,
\end{eqnarray}
of which the last term is a positive real number. Hence using (\ref{fA}) and (\ref{fB}), the inequality (\ref{Sch2}) reduces to
\begin{eqnarray}
\label{Sch3}
\langle (\Delta A(t_1))^2 \rangle ~ \langle (\Delta B(t_2))^2 \rangle \ge \frac{1}{4} \Big| \langle [A(t_1), B(t_2)] \rangle \Big|^2.
\end{eqnarray}
Taking square root on both sides of (\ref{Sch3}), we establish the temporal uncertainty relation (\ref{hupT}). Let us now
demonstrate the dynamical Heisenberg-Robertson uncertainty relation (\ref{hupT}) for real physical observables that are measured
at two different times. First we consider a spin-1/2 quantum system evolving under a particular Hamiltonian ${\mathcal H} = \omega S_z$.
We consider the measurement of spin observables $S_x$ and $S_y$ at two different times $t_1$ and $t_2$. The time evolution
of the spin operators in the Heisenberg picture are given by
\begin{eqnarray}
\label{Sx}
& \frac{d}{dt} S_x(t) = \frac{1}{i\hbar} \left[S_x(t),  {\mathcal H} \right], \\
\label{Sy}
& \frac{d}{dt} S_y(t) = \frac{1}{i\hbar} \left[S_y(t),  {\mathcal H} \right],
\end{eqnarray}
where $S_x(t)$ and $S_y(t)$ are the spin operators in Heisenberg picture given by
\begin{eqnarray}
\label{SxH}
&S_x(t) = e^{i {\mathcal H} t / \hbar} ~ S_x ~ e^{ - i {\mathcal H} t / \hbar}, \\
\label{SyH}
&S_y(t) = e^{i {\mathcal H} t / \hbar} ~ S_y ~ e^{ - i {\mathcal H} t / \hbar}.
\end{eqnarray}
The spin operators
\begin{eqnarray}
\nonumber
S_{x} = \frac{\hbar}{2} \left(\begin{array}{cc}
 0  & 1\\
1 & 0\\
\end{array}
\right),
S_{y} = \frac{\hbar}{2} \left(\begin{array}{cc}
 0  & -i\\
i & 0\\
\end{array}
\right),
S_{z} = \frac{\hbar}{2} \left(\begin{array}{cc}
 1  & 0\\
0 & -1\\
\end{array}
\right)
\end{eqnarray}
satisfy the commutation relations $[S_{x},S_{y}]=i\hbar S_{z}$, $[S_{y},S_{z}]$$=i\hbar S_{x}$,
and $[S_{z},S_{x}]$$=i\hbar S_{y}$. Then using the commutation relations $[S_{x},{\mathcal H}]$$=-i \hbar \omega S_{y}$
and $[S_{y},{\mathcal H}]$$=i \hbar \omega S_{x}$, one can obtain Heisenberg equations of motion (\ref{Sx}) and
(\ref{Sy}) for $S_x(t)$ and $S_y(t)$ as
\begin{eqnarray}
\label{SxSy}
\frac{d}{dt} S_x(t) = -\omega S_y(t), ~\mbox{and}~ \frac{d}{dt} S_y(t) = \omega S_x(t).
\end{eqnarray}
Solutions of the Heisenberg equations gives us the time evolution of the spin operators
\begin{eqnarray}
\label{SxSol}
& S_x(t) = S_x \cos(\omega t) - S_y \sin(\omega t), \\
\label{SySol}
& S_y(t) = S_y \cos(\omega t) + S_x \sin(\omega t).
\end{eqnarray}
Considering the measurement of spin observables $S_x$ and $S_y$ at two different times $t_1$ and $t_2$, we can
write down the temporal version of the Heisenberg-Robertson uncertainty relation (\ref{hupT}) as follows
\begin{eqnarray}
\label{hupSxSy1}
\Delta S_x(t_1) \Delta S_y(t_2) \ge \frac{1}{2} \Big| \langle [S_x(t_1),S_y(t_2)] \rangle \Big|,
\end{eqnarray}
where the measurement uncertainties
\begin{eqnarray}
\label{dSxT1}
& \Delta S_x(t_1)=\sqrt{\langle S^2_x(t_1)\rangle - \langle S_x(t_1)\rangle^2}, \\
\label{dSyT2}
& \Delta S_y(t_2)=\sqrt{\langle S^2_y(t_2)\rangle - \langle S_y(t_2)\rangle^2}
\end{eqnarray}
are restricted by the commutator
\begin{eqnarray}
\label{Cx1y2}
\left[S_x(t_1), S_y(t_2) \right] = i \hbar \cos [\omega (t_2 - t_1)] S_z,
\end{eqnarray}
and the uncertainty relation (\ref{hupSxSy1}) takes the form
\begin{eqnarray}
\label{hupSxSy2}
\Delta S_x(t_1) \Delta S_y(t_2) \ge \frac{\hbar}{2} \Big| \cos [\omega (t_2 - t_1)] \langle S_z \rangle \Big|.
\end{eqnarray}
Hence the measurement uncertainties are now correlated in time. The uncertainties not only depend on the choice
of observables, they also depend on the time at which the physical observables are measured. The commutator
$\left[S_x(t_1), S_y(t_2) \right]$ dictates the trade-off between the dynamical uncertainties. Now, under the
limiting situation when $t_1$$=t_2$$=t$, the two-time uncertainty relation (\ref{hupSxSy2}) reduces to the
well known equal time uncertainty relation
\begin{eqnarray}
\label{hupSxSy3}
\Delta S_x(t) \Delta S_y(t) \ge \frac{\hbar}{2} \Big| \langle S_z \rangle \Big|.
\end{eqnarray}
We can also interchange the sequence of measurements. Suppose we first measure $S_y$ at $t_1$
and then measure $S_x$ at a later time $t_2$. The uncertainty relation in that case becomes
\begin{eqnarray}
\label{hupSySx1}
\Delta S_y(t_1) \Delta S_x(t_2) \ge \frac{1}{2} \Big| \langle [S_y(t_1),S_x(t_2)] \rangle \Big|,
\end{eqnarray}
where the measurement uncertainties
\begin{eqnarray}
\label{dSyT1}
&\Delta S_y(t_1)=\sqrt{\langle S^2_y(t_1)\rangle - \langle S_y(t_1)\rangle^2}, \\
\label{dSxT2}
&\Delta S_x(t_2)=\sqrt{\langle S^2_x(t_2)\rangle - \langle S_x(t_2)\rangle^2}
\end{eqnarray}
are restricted by the commutator
\begin{eqnarray}
\label{Cy1x2}
\left[S_y(t_1), S_x(t_2) \right] = -i \hbar \cos [\omega (t_2 - t_1)] S_z,
\end{eqnarray}
and the uncertainty relation (\ref{hupSySx1}) takes the form
\begin{eqnarray}
\label{hupSySx2}
\Delta S_y(t_1) \Delta S_x(t_2) \ge \frac{\hbar}{2} \Big| \cos [\omega (t_2 - t_1)] \langle S_z \rangle \Big|.
\end{eqnarray}
Next, we present the uncertainty relation for the same spin observable $S_x$ measured at two different
times $t_1$ and $t_2$. Following the general form of the Heisenberg-Robertson uncertainty relation
(\ref{hupT}) we have in this case
\begin{eqnarray}
\label{hupSxSx1}
\Delta S_x(t_1) \Delta S_x(t_2) \ge \frac{1}{2} \Big| \langle [S_x(t_1),S_x(t_2)] \rangle \Big|.
\end{eqnarray}
The commutator on the right hand side of (\ref{hupSxSx1}) can be evaluated using the time-evolved
spin operator (\ref{SxSol}) as
\begin{eqnarray}
\label{Cx1x2}
\left[S_x(t_1), S_x(t_2) \right] = -i \hbar \sin [\omega (t_2 - t_1)] S_z,
\end{eqnarray}
resulting to the following uncertainty relation
\begin{eqnarray}
\label{hupSxSx2}
\Delta S_x(t_1) \Delta S_x(t_2) \ge \frac{\hbar}{2} \Big| \sin [\omega (t_2 - t_1)] \langle S_z \rangle \Big|.
\end{eqnarray}
If the two successive measurements are very close in time, then ($t_2 - t_1$)$\rightarrow 0$ and
$\sin [\omega (t_2 - t_1)]$ $\sim$ $\omega (t_2 - t_1)$. The dynamical uncertainty relation (\ref{hupSxSx2})
in this situation reduces to
\begin{eqnarray}
\label{hupSxSx3}
\Delta S_x(t_1) \Delta S_x(t_2) \ge \frac{\hbar}{2} \Big| \omega (t_2 - t_1) \langle S_z \rangle \Big|,
\end{eqnarray}
reflecting the fact that when one measures the spin observable $S_x$ successively at two different
times $t_1$ and $t_2$ with ($t_2 - t_1$)$\rightarrow 0$, the uncertainty of the second measurement
$\Delta S_x(t_2)$ will be very small. Similarly, the uncertainty relation corresponding to the measurement
of spin component $S_y$ at two different times $t_1$ and $t_2$ is given by
\begin{eqnarray}
\label{hupSySy1}
\Delta S_y(t_1) \Delta S_y(t_2) \ge \frac{1}{2} \Big| \langle [S_y(t_1),S_y(t_2)] \rangle \Big|.
\end{eqnarray}
The right hand side commutator of (\ref{hupSySy1}) can be evaluated again using the time-evolved
spin operator (\ref{SySol}) as
\begin{eqnarray}
\label{Cy1y2}
\left[S_y(t_1), S_y(t_2) \right] = -i \hbar \sin [\omega (t_2 - t_1)] S_z,
\end{eqnarray}
resulting to the following uncertainty relation
\begin{eqnarray}
\label{hupSySy2}
\Delta S_y(t_1) \Delta S_y(t_2) \ge \frac{\hbar}{2} \Big| \sin [\omega (t_2 - t_1)] \langle S_z \rangle \Big|.
\end{eqnarray}
To verify the time-dependent uncertainty relations, one can prepare the spin-1/2 system in some quantum state
\begin{eqnarray}
| \Psi_s \rangle = \cos (\theta/2) | \uparrow \rangle + \sin (\theta/2) | \downarrow \rangle,
\end{eqnarray}
for which the expectation value
\begin{eqnarray}
\langle S_z \rangle = \langle \Psi_s | S_z | \Psi_s \rangle=(\hbar/2)\cos (\theta).
\end{eqnarray}
Using Eqs.~(\ref{dSxT1}), (\ref{dSyT2}), (\ref{dSyT1}), and (\ref{dSxT2}) one can obtain the time-dependent
uncertainties quantified by the variances with respect to the state $| \Psi_s \rangle$ as given below
\begin{eqnarray}
\label{deltaSxT1}
\Delta S_x(t_1) = \frac{\hbar}{2} \sqrt{1-\sin^2(\theta) \cos^2 (\omega t_1)},
\end{eqnarray}
\vskip -0.7cm
\begin{eqnarray}
\label{deltaSyT2}
\Delta S_y(t_2) = \frac{\hbar}{2} \sqrt{1-\sin^2(\theta) \sin^2 (\omega t_2)},
\end{eqnarray}
\vskip -0.7cm
\begin{eqnarray}
\label{deltaSyT1}
\Delta S_y(t_1) = \frac{\hbar}{2} \sqrt{1-\sin^2(\theta) \sin^2 (\omega t_1)},
\end{eqnarray}
\vskip -0.7cm
\begin{eqnarray}
\label{deltaSxT2}
\Delta S_x(t_2) = \frac{\hbar}{2} \sqrt{1-\sin^2(\theta) \cos^2 (\omega t_2)}.
\end{eqnarray}
Explicit forms of the time dependent quantum uncertainty relations can be obtained by substituting these expressions
of $\langle S_z \rangle$, $\Delta S_x(t_1)$, $\Delta S_y(t_2)$, $\Delta S_y(t_1)$, and $\Delta S_x(t_2)$
into the equations (\ref{hupSxSy2}), (\ref{hupSySx2}), (\ref{hupSxSx2}), and (\ref{hupSySy2}). The resulting temporal
uncertainty relations can be experimentally verified with the present superconducting quantum technology \cite{SCQ}
or with trapped ion qubits \cite{Blatt08,Kirchmair09}.

Next, we explore the temporal version of the Heisenberg-Robertson uncertainty relation (\ref{hupT}) for a quantum
harmonic oscillator evolving under the Hamiltonian
\begin{eqnarray}
{\mathscr H} = \frac{P^2}{2m} + \frac{1}{2} m \omega^2 X^2
\end{eqnarray}
where $X$ and $P$ are the position and momentum operators satisfying the commutation relation $\left[X,P\right]=i\hbar$.
In the Heisenberg picture, the operators $X$ and $P$ evolves with time according to the equations
\begin{eqnarray}
\label{X}
\frac{d}{dt} X(t) = \frac{1}{i\hbar} \left[X(t),  {\mathscr H} \right],
\end{eqnarray}
\vskip -0.5cm
\begin{eqnarray}
\label{P}
\frac{d}{dt} P(t) = \frac{1}{i\hbar} \left[P(t),  {\mathscr H} \right],
\end{eqnarray}
where $X(t)$ and $P(t)$ are position and momentum operators in the Heisenberg picture given by
\begin{eqnarray}
\label{XH}
&X(t) = e^{i {\mathscr H} t / \hbar} ~ X ~ e^{ - i {\mathscr H} t / \hbar}, \\
\label{PH}
&P(t) = e^{i {\mathscr H} t / \hbar} ~ P ~ e^{ - i {\mathscr H} t / \hbar}.
\end{eqnarray}
Using the commutation relations $[X, {\mathscr H}]$$= (i \hbar/m) P$ and $[P, {\mathscr H}]$$=(-i \hbar m \omega^2) X$,
one can solve the Heisenberg equations of motion (\ref{X}) and (\ref{P}) to obtain the exact time evolution
of the operators as
\begin{eqnarray}
\label{XSol}
X(t) = X \cos(\omega t) + \frac{1}{m \omega} P \sin(\omega t),
\end{eqnarray}
\vskip -0.7cm
\begin{eqnarray}
\label{PSol}
P(t) = P \cos(\omega t) - m \omega X \sin(\omega t).
\end{eqnarray}
The temporal version of Heisenberg-Robertson uncertainty relation (\ref{hupT}) for measurement of observables
$X$ and $P$ at two different times $t_1$ and $t_2$ is given by
\begin{eqnarray}
\label{hupXP1}
\Delta X(t_1) \Delta P(t_2) \ge \frac{1}{2} \Big| \langle [X(t_1),P(t_2)] \rangle \Big|,
\end{eqnarray}
where the measurement uncertainties $\Delta X(t_1)$, $\Delta P(t_2)$, and the commutator on the right
can easily be obtained using the time evolved operators in (\ref{XSol}) and (\ref{PSol})
\begin{eqnarray}
\label{dXT1}
& \Delta X(t_1)=\sqrt{\langle X^2(t_1)\rangle - \langle X(t_1)\rangle^2}, \\
\label{dPT2}
& \Delta P(t_2)=\sqrt{\langle P^2(t_2)\rangle - \langle P(t_2)\rangle^2}, \\
\label{CX1P2}
& \left[X(t_1), P(t_2) \right] = i \hbar \cos [\omega (t_2 - t_1)],
\end{eqnarray}
that leads the uncertainty relation (\ref{hupXP1}) to the form
\begin{eqnarray}
\label{hupXP2}
\Delta X(t_1) \Delta P(t_2) \ge \frac{\hbar}{2} \Big| \cos [\omega (t_2 - t_1)] \Big|.
\end{eqnarray}
The measurement uncertainties are again restricted by the time-correlated commutator. Under a
situation when $t_1=t_2=t$, the two-time uncertainty relation (\ref{hupXP2}) reduces to the
equal time uncertainty relation
\begin{eqnarray}
\label{hupXP3}
\Delta X(t) \Delta P(t) \ge \frac{\hbar}{2}.
\end{eqnarray}
Moreover, we can interchange the order of measurements. We can first measure momentum $P$ at $t_1$
and then measure the position $X$ at a later time $t_2$. Then the form of the uncertainty relation becomes
\begin{eqnarray}
\label{hupPX1}
\Delta P(t_1) \Delta X(t_2) \ge \frac{1}{2} \Big| \langle [P(t_1),X(t_2)] \rangle \Big|.
\end{eqnarray}
where the measurement uncertainties $\Delta P(t_1)$, $\Delta X(t_2)$, and the two-time commutator
on the right can be obtained using the time-evolved position and momentum operators in (\ref{XSol})
and (\ref{PSol}) as
\begin{eqnarray}
\label{dPT1}
& \Delta P(t_1)=\sqrt{\langle P^2(t_1)\rangle - \langle P(t_1)\rangle^2}, \\
\label{dXT2}
& \Delta X(t_2)=\sqrt{\langle X^2(t_2)\rangle - \langle X(t_2)\rangle^2}, \\
\label{CP1X2}
& \left[P(t_1), X(t_2) \right] = -i \hbar \cos [\omega (t_2 - t_1)],
\end{eqnarray}
consequently the uncertainty relation (\ref{hupPX1}) takes the form
\begin{eqnarray}
\label{hupPX2}
\Delta P(t_1) \Delta X(t_2) \ge \frac{\hbar}{2} \Big| \cos [\omega (t_2 - t_1)] \Big|.
\end{eqnarray}
We now consider the measurement of position $X$ at two different times $t_1$ and $t_2$.
The Heisenberg-Robertson uncertainty relation (\ref{hupT}) in this case becomes
\begin{eqnarray}
\label{hupXX1}
\Delta X(t_1) \Delta X(t_2) \ge \frac{1}{2} \Big| \langle [X(t_1), X(t_2)] \rangle \Big|.
\end{eqnarray}
The two-time commutator on the right hand side of (\ref{hupXX1}) can be evaluated using the
time evolution of the position operator (\ref{XSol}) as
\begin{eqnarray}
\label{DX1X2}
\left[X(t_1), X(t_2) \right] = \frac{i \hbar}{m \omega} \sin [\omega (t_2 - t_1)],
\end{eqnarray}
resulting to the dynamical uncertainty relation
\begin{eqnarray}
\label{hupXX2}
\Delta X(t_1) \Delta X(t_2) \ge \frac{\hbar}{2 m \omega} \Big| \sin [\omega (t_2 - t_1)] \Big|.
\end{eqnarray}
Consider the two successive measurements are very close in time, then
($t_2 - t_1$)$\rightarrow 0$ and $\sin [\omega (t_2 - t_1)]$ $\sim$ $\omega (t_2 - t_1)$.
The uncertainty relation (\ref{hupXX2}) under that situation reduces to
\begin{eqnarray}
\label{hupXX3}
\Delta X(t_1) \Delta X(t_2) \ge \frac{\hbar}{2 m} \Big| t_2 - t_1 \Big|,
\end{eqnarray}
which implies that if one measures the observable $X$ successively at two different
times $t_1$ and $t_2$ with ($t_2 - t_1$)$\rightarrow 0$, the uncertainty of the second measurement
$\Delta X(t_2)$ will be negligibly small. Furthermore, the temporal uncertainty relation corresponding
to the measurement of momentum operator $P$ at two different times $t_1$ and $t_2$ is given by
\begin{eqnarray}
\label{hupPP1}
\Delta P(t_1) \Delta P(t_2) \ge \frac{1}{2} \Big| \langle [P(t_1),P(t_2)] \rangle \Big|,
\end{eqnarray}
where the commutator in the right hand side of (\ref{hupPP1}) can be evaluated again using
the time-evolved momentum operator (\ref{PSol}) as
\begin{eqnarray}
\label{DP1P2}
\left[P(t_1), P(t_2) \right] = i \hbar m \omega \sin [\omega (t_2 - t_1)],
\end{eqnarray}
that leads to the following uncertainty relation
\begin{eqnarray}
\label{hupPP2}
\Delta P(t_1) \Delta P(t_2) \ge \frac{\hbar m \omega}{2} \Big| \sin [\omega (t_2 - t_1)] \Big|.
\end{eqnarray}
For experimental verification of the time-dependent uncertainty relations, one can prepare the
system in a superposition of energy eigenstates given by
\begin{eqnarray}
| \Psi_h \rangle = \frac{1}{\sqrt{2}} \left( | 0 \rangle + | 1 \rangle \right)
\end{eqnarray}
where $| 0 \rangle$ and $| 1 \rangle$ are the ground state and the first excited state of the quantum harmonic
oscillator. Finally, using the equations ~(\ref{dXT1}), (\ref{dPT2}), (\ref{dPT1}), and (\ref{dXT2}) one can
obtain various time-dependent uncertainties quantified by the variances with respect to this state $| \Psi_h \rangle$
as follows
\begin{eqnarray}
\label{deltaXT1}
\Delta X(t_1) = \sqrt{\frac{\hbar}{2m\omega} \big(2-\cos^2 (\omega t_1)  \big)},
\end{eqnarray}
\vskip -0.7cm
\begin{eqnarray}
\label{deltaPT2}
\Delta P(t_2) = \sqrt{\frac{m\hbar\omega}{2} \big(2-\sin^2 (\omega t_2) \big)},
\end{eqnarray}
\vskip -0.7cm
\begin{eqnarray}
\label{deltaPT1}
\Delta P(t_1) = \sqrt{\frac{m\hbar\omega}{2} \big( 2-\sin^2 (\omega t_1) \big)},
\end{eqnarray}
\vskip -0.7cm
\begin{eqnarray}
\label{deltaXT2}
\Delta X(t_2) = \sqrt{\frac{\hbar}{2m\omega} \big( 2-\cos^2 (\omega t_2) \big)}.
\end{eqnarray}
One can substitute these expressions of $\Delta X(t_1)$, $\Delta P(t_2)$, $\Delta P(t_1)$,
and $\Delta X(t_2)$ in equations (\ref{hupXP2}), (\ref{hupPX2}), (\ref{hupXX2}), and (\ref{hupPP2})
to obtain the explicit forms of the time-correlated quantum uncertainty relations. The resulting dynamical
uncertainty relations can be experimentally verified using for example, nanomechanical oscillator \cite{NanoMech}, superconducting
quantum circuit \cite{Martinis08}, or quantum optical field \cite{Haroche11}. It will also be interesting to investigate
the time-dependent uncertainty relation (\ref{hupT}) for a wide range of quantum mechanical problems.

\begin{acknowledgments}
Md.~Manirul Ali was supported by the Centre for Quantum Science and Technology, Chennai Institute of
Technology, India, vide funding number CIT/CQST/2021/RD-007.
\end{acknowledgments}

\end{document}